\documentclass[runningheads]{llncs}
\usepackage[T1]{fontenc}
\usepackage{graphicx}
\usepackage[backref=page]{hyperref}
\usepackage{xurl} 
\usepackage{orcidlink}
\usepackage{xcolor}
\hypersetup{
    colorlinks=true,
    linkcolor=black,
    urlcolor=blue,
    citecolor=black,
    filecolor=black,
    linkbordercolor=white, 
    urlbordercolor=white,
    citebordercolor=white,
    filebordercolor=white,
}

\usepackage{hhline}
\usepackage{multirow}
\usepackage{makecell}
\usepackage{comment}

\usepackage{siunitx}

\usepackage{float} 
\usepackage[noabbrev]{cleveref}

\usepackage{color}

\usepackage[misc]{ifsym}

\begin{document}

\title{
\texorpdfstring{
Data-Centric Strategies for Overcoming\\
PET/CT Heterogeneity: Insights from the AutoPET III Lesion Segmentation Challenge}
{Data-Centric Strategies for Overcoming PET/CT Heterogeneity: Insights from the AutoPET III Lesion Segmentation Challenge}
}

%
%
\author{
        Balint Kovacs\inst{1,2,\text{\Letter},\star}\orcidlink{0000-0002-1191-0646}\and
        Shuhan Xiao\inst{1,3,\text{\Letter},\star}\orcidlink{0000-0001-5397-814X}\and
        Maximilian Rokuss\inst{1,3}\orcidlink{0009-0004-4560-0760}\and\\
        Constantin Ulrich\inst{1,2}\orcidlink{0000-0003-3002-8170}\and
        Fabian Isensee\inst{1,4,\mu}\orcidlink{0000-0002-3519-5886}\and
        Klaus H. Maier-Hein\inst{1,4,5,\mu}\orcidlink{0000-0002-6626-2463}
}
\authorrunning{B. Kovacs and S. Xiao et al.}
\titlerunning{Adapting to the Variability in PET/CT Imaging}

\institute{
German Cancer Research Center (DKFZ) Heidelberg,\\Division of Medical Image Computing, Heidelberg, Germany
\and
Medical Faculty Heidelberg, Heidelberg University, Heidelberg, Germany\and
Faculty of Mathematics and Computer Science,\\Heidelberg University, Heidelberg, Germany\and
Helmholtz Imaging, DKFZ, Heidelberg, Germany\and
Pattern Analysis and Learning Group, Department of Radiation Oncology, Heidelberg University Hospital, Heidelberg, Germany
\\\email{\{balint.kovacs, s.xiao\}@dkfz-heidelberg.de}
}

\maketitle              
\renewcommand{\thefootnote}{$\star$}
\footnotetext[1]{Equal contribution}
\renewcommand{\thefootnote}{$\mu$}
\footnotetext[2]{Shared last authorship}

\begin{abstract}
The third autoPET challenge introduced a new data-centric task this year, shifting the focus from model development to improving metastatic lesion segmentation on PET/CT images through data quality and handling strategies. In response, we developed targeted strategies to enhance segmentation performance tailored to the characteristics of PET/CT imaging. Our approach encompasses two key elements. First, to address potential misalignments between CT and PET modalities as well as the prevalence of punctate lesions, we modified the baseline data augmentation scheme and extended it with misalignment augmentation. This adaptation aims to improve segmentation accuracy, particularly for tiny metastatic lesions. Second, to tackle the variability in image dimensions significantly affecting the prediction time, we implemented a dynamic ensembling and test-time augmentation (TTA) strategy. This method optimizes the use of ensembling and TTA within a 5-minute prediction time limit, effectively leveraging the generalization potential for both small and large images. Both of our solutions are designed to be robust across different tracers and institutional settings, offering a general, yet imaging-specific approach to the multi-tracer and multi-institutional challenges of the competition. We made the challenge repository with our modifications publicly available at \url{https://github.com/MIC-DKFZ/miccai2024_autopet3_datacentric}.

\end{abstract}

\section{Introduction}
Accurate automated tumor lesion segmentation from whole-body PET/CT scans has become increasingly essential in oncological diagnostics, supporting tumor characterization, treatment planning, and response monitoring \cite{foster2014review}. The international AutoPET competition, now in its third year, is aimed at improving this task using artificial intelligence (AI) \cite{ingrisch_2024_10990932}. While the general goal of the challenge remains to achieve the most accurate lesion segmentation with a strong emphasis on generalizability using a multi-institutional test set, the 2024 challenge edition introduces a new focus on cross-center and tracer generalizability. The publicly available dataset of 1014 FDG PET/CT studies \cite{gatidis2020fdgpetct} has been extended by 597 exams with a new PSMA tracer \cite{jeblick2024psmapetct}, alongside an extended hidden test dataset, prompting participants to develop even more robust AI solutions.

Motivated by the inherent complexity and imperfections of real-world data, a new data-centric task has been also introduced. Unlike traditional model-centric approaches, where performance gains are often sought by building larger and more complex models, this task focuses on improving data quality and handling, such as image pre- and post-processing, and data augmentation. In this case, the model architecture is fixed and cannot be modified. Data-centric AI, an emerging field, demonstrates that in many cases, enhancing the quality of the data is the most effective way to improve performance in practical machine learning applications \cite{singh2023systematic,zha2023data}.

Though PET and CT images are acquired from the same machine during the same examination, misalignments between the modalities can occur due to patient movements. Involuntary organ movements can be further exaggerated by attenuation correction or imaging artifacts during the imaging process \cite{alessio2004pet,hunter2016patient,kaji2024improvement,lodge2011effect}. These misalignments can result in inconsistencies in the ground truth between image modalities and may be particularly critical when dealing with tiny metastatic lesions prominent in the dataset. Since semantic segmentation networks rely heavily on spatial information, such inconsistencies can potentially limit segmentation performance.

The addition of PSMA cases to the autoPET III training dataset has significantly increased the already high variability in geometrical image properties. The dataset includes scans with varying dimensions, ranging from smaller, regional scans (mid-femur to neck) to full-body scans (feet to head), as well as differences in image spacing, further contributing to variability. These factors lead to substantially varying prediction times between cases. To ensure that all scans, especially the larger ones, meet the 5-minute inference time limit, a fixed ensembling and test-time augmentation (TTA) strategy would require setting conservative parameters, which would significantly limit their generalization ability for smaller images.

In this paper, we present our solutions to the challenges outlined above. To address potential misalignments between the image sequences and the prevalence of punctate lesions in the dataset, we modified and extended the data augmentation (DA) scheme of the baseline model provided by the challenge organizers, incorporating misalignment augmentation \cite{kovacs2023addressing}. To fully exploit the generalization ability of ensembling and TTA for all cases with different dimensions, we implemented a dynamic prediction strategy by optimizing the extent of ensembling and the number of TTA, while still ensuring timely scan processing. These general, yet imaging-specific approaches are independent of tracer type and institutional device settings, offering a potentially robust solution for the multi-tracer, multi-institutional nature of the challenge.

\section{Methods}
\subsection{Challenge Design - The Data-Centric Task}
This work proposes solutions for the data-centric task (Category 2) of the challenge. The challenge organizers provided a framework (\url{https://github.com/ClinicalDataScience/datacentric-challenge/tree/main}) for this task, which is based on the nnU-Net ~\cite{isensee2021nnu} training configuration and to which no modifications are allowed.

\subsection{Data augmentation strategy}\label{subsec:DA}
We adapt the baseline DA scheme to be more suitable for our image modalities PET and CT. We perform less aggressive and more realistic random affine transformations by reducing all amplitudes of its hyperparameters. While the baseline DA scheme adds random Gaussian smoothing, we remove this transformation to avoid blurring small lesions and to preserve fine image details. To prevent overexposure of the images leading to information loss, we also remove the contrast transforms that invert the images before performing gamma transformations and re-inverting them back after. We refer to this setting as \textit{subtleDA}. 
    \begin{figure}[H]
		\begin{center}
			\includegraphics[width=0.7\textwidth]{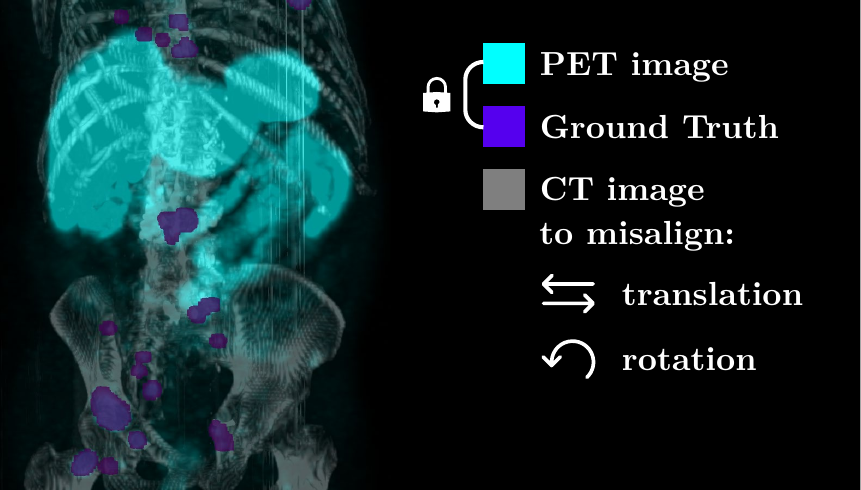}{}
			\caption{Misalignment augmentation applied to the dataset. The PET image remains unchanged, while the CT image is transformed to introduce additional slight plausible misalignments between the image modalities. The ground truth is coupled with the PET modality due to its stronger relevance for identifying metastatic lesions.
            }
			\label{fig:misalDA} {}
		\end{center} {}
	\end{figure}

Further, to address potential misalignments between the CT and PET images, we extended the DA scheme with misalignment DA (\textit{misalDA})~\cite{kovacs2023addressing}. This approach has the potential advantage of improving sensitivity for punctate lesions with small voxel segmentation volumes, which was suggested, though not proven, in the original study. Given the prevalence of such lesions in our dataset, the segmentation of such punctate lesions plays a particularly crucial role. The amplitude of the transformations used to generate misalignments was sampled randomly from a uniform distribution, constrained by a maximum amplitude in both positive and negative directions. The transformation included an initial rotation with a maximum angle of \SI{5}{\degree}, followed by translations with maximum voxel shifts of [2, 2, 0] in the x, y, and z directions, respectively. Both transformations were applied with \SI{10}{\percent} probability. Since the PET modality contains the most important information regarding metastatic lesions and correlates more closely with the ground truth than the CT modality, only the CT images were displaced, while the PET and ground truth images remained unchanged after the transformation (see \cref{fig:misalDA}).
\par We hypothesize that the large prevalence of small metastatic lesions can be effectively addressed by preserving fine image details using the subtleDA scheme and meanwhile increasing the sensitivity to punctate lesions using misalDA.

To achieve maximum variability in the training data, we opt for online data augmentation rather than offline augmentation, where pre-computed augmented images are saved. 

\subsection{Network training}\label{subsec:training}
We perform 5-fold cross-validation and train 3D U-Net models on the dataset according to the fixed architecture and training scheme baseline for 250k training steps. We train the network with both the baseline and subtleDA scheme with and without misalDA (see section 2.2.).

\subsection{Image segmentation postprocessing}
For the postprocessing step following our predictions, we mask the predicted segmentation regions in areas where the standard uptake value (SUV) in the PET images was lower than 1.0 to reduce false positive volumes, similarly to the baseline framework.

\subsection{Dynamic ensembling and test-time augmentation} \label{subsec:ensembling}
Due to the time limit of \SI{5}{\min} per case set by the challenge, we constrain the extent of model ensembling and the number of TTAs that can be applied. In many cases, a full 5-fold ensembling with multiple TTAs is feasible within this time limit. However, for samples that require resampling or have large image sizes, the maximum number of ensembled models is reduced to 2-3, often without any TTA. To stay within the time limit while maintaining the benefits of both ensembling and TTA, we implemented a case-specific dynamic ensembling approach in addition to the dynamic TTA approach in the baseline framework. Specifically, we set a maximum time of \SI{25}{\sec} per model for TTA and \SI{170}{\sec} for the whole ensembling including TTA. In the dynamic TTA approach, the initial prediction by each model, the number of TTAs that can be applied within the given time constraint (\SI{25}{\sec}) is determined by the time of the first inference pass, but not more than a predefined maximum number of TTA of 2. Similarly, we approximate the maximum number of models that can be ensembled within the remaining ensemble time limit based on the time it takes for the first model including the possible TTAs to finish its predictions.

\subsection{Evaluation} \label{subsec:evaluation}
The mean Dice score of the segmented lesions (\SI{50}{\percent} contribution), false positive volume (FPvol) with \SI{25}{\percent} contribution, and false negative volume (FNvol) with \SI{25}{\percent} contribution will be used as metrics in the final evaluation of the challenge. Given that the test set consists of \SI{50}{\percent} FDG and \SI{50}{\percent} PSMA exams, we additionally focus on the balanced (bal.) Dice score, weighted by the number of exams from each tracer, to simulate the tracer distribution of \SI{50}{\percent} of the data in the hidden test set. Since no separate development test set is available in the challenge, we are unable to fully validate the effectiveness of the dynamic ensembling and TTA strategy. Therefore, we evaluate all of our trained models using all mirroring transformations as part of the TTA on our validation folds.

\section{Results}
We systematically summarize our results in \cref{table:resultsDC}. Using the subtleDA scheme alone or extending the baseline DA scheme with misalDA does not lead to an improvement with respect to the mean or balanced Dice scores. However, combining the subtle DA scheme with misalDA resulted in the best Dice score with an improvement of 0.81 dice points over the baseline and a decrease of average FPvol by about \SI{6}{\milli\liter}.
   \begin{table}[H]
        \begin{center}
        \caption{Five-fold cross-validation results for our data-centric solutions showing the mean Dice score as well as the FP and FN volume in \si{\milli\liter}.}
        \label{table:resultsDC}
            \begin{tabular}{
            p{0.35\textwidth}<{\raggedright}
            >{\centering\arraybackslash}p{0.09\textwidth}
            >{\centering\arraybackslash}p{0.09\textwidth}
            >{\centering\arraybackslash}p{0.09\textwidth}
            >{\centering\arraybackslash}p{0.09\textwidth}
            >{\centering\arraybackslash}p{0.09\textwidth}
            >{\centering\arraybackslash}p{0.09\textwidth}
            }
        
        & \multicolumn{4}{c}{Dice$\uparrow$} \\  
        \cline{2-5}
        DA scheme & FDG   & PSMA  & mean  & bal.     &FPvol$\downarrow$ & FNvol$\downarrow$ \\ 
        \hline
        baseline  & 57.64 & 49.05 & 53.27 & 52.23    & 6.09             & 30.90             \\
        baseline+misalDA  & 54.09 & 47.64 & 50.76    & 50.03 & \textbf{5.69} & 33.37 \\
        subtleDA  & 55.26 & 47.65 & 51.36 & 50.47    & 6.74             & 29.82             \\
        subtleDA+misalDA & 57.50 & 50.92 & \textbf{54.08} & \textbf{53.36} & 8.37 & \textbf{24.75}
        \end{tabular}
        \end{center}
    \end{table}

\newpage
\section{Discussion and conclusion}
Considering the distribution shift among the public development set and hidden test set, our contribution to the challenge aimed to leverage the potential of heterogeneity in PET/CT imaging to provide robust predictions accounting for multi-modal misalignments and geometrical image properties.

As no separate validation set is available, the true impact of our dynamic ensembling and TTA strategy will only be revealed once the results from the hidden challenge test set are made public.

Removing transformations from the baseline DA scheme that could potentially eliminate the detection of punctate metastatic lesions and meanwhile extending it with misalignment augmentation led us to the settings we used for our final solution. This result indicates the success of our strategy to be sensitive for segmenting tiny metastatic lesions, but its confirmation needs further detailed analysis. The decrease in performance for the baseline DA scheme with misalDA highlights the importance of the transformation combination to leverage their full synergies. 

Overall, our approach demonstrates the importance of data augmentation to improve segmentation accuracy, though further analysis is needed to fully validate these findings.

\begin{credits}
\subsubsection{\ackname}
Part of this work was funded by Helmholtz Imaging (HI), a platform of the Helmholtz Incubator on Information and Data Science.
\end{credits}

\newpage
\bibliographystyle{splncs04}
\bibliography{biblio2}

\begin{thebibliography}{10}
\providecommand{\url}[1]{\texttt{#1}}
\providecommand{\urlprefix}{URL }
\providecommand{\doi}[1]{https://doi.org/#1}

\bibitem{alessio2004pet}
Alessio, A.M., Kinahan, P.E., Cheng, P.M., Vesselle, H., Karp, J.S.: {PET}/{CT} scanner instrumentation, challenges, and solutions. Radiologic Clinics  \textbf{42}(6),  1017--1032 (2004). \doi{doi:10.1016/j.rcl.2004.08.001}

\bibitem{foster2014review}
Foster, B., Bagci, U., Mansoor, A., Xu, Z., Mollura, D.J.: A review on segmentation of positron emission tomography images. Computers in biology and medicine  \textbf{50},  76--96 (2014)

\bibitem{gatidis2020fdgpetct}
Gatidis, S., Kuestner, T.: A whole-body fdg-pet/ct dataset with manually annotated tumor lesions (fdg-pet-ct-lesions). \url{https://doi.org/10.7937/GKR0-XV29} (2022), the Cancer Imaging Archive

\bibitem{hunter2016patient}
Hunter, C.R., Klein, R., Beanlands, R.S., deKemp, R.A.: Patient motion effects on the quantification of regional myocardial blood flow with dynamic {PET} imaging. Medical physics  \textbf{43}(4),  1829--1840 (2016). \doi{DOI: 10.1118/1.4943565}

\bibitem{ingrisch_2024_10990932}
Ingrisch, M., Dexl, J., Jeblick, K., Cyran, C., Gatidis, S., Kuestner, T.: {Automated Lesion Segmentation in Whole-Body PET/CT - Multitracer Multicenter generalization} (Apr 2024). \doi{10.5281/zenodo.10990932}

\bibitem{isensee2021nnu}
Isensee, F., Jaeger, P.F., Kohl, S.A., Petersen, J., Maier-Hein, K.H.: nnu-net: a self-configuring method for deep learning-based biomedical image segmentation. Nature methods  \textbf{18}(2),  203--211 (2021). \doi{https://doi.org/10.1038/s41592-020-01008-z}

\bibitem{jeblick2024psmapetct}
Jeblick, K., et~al.: A whole-body psma-pet/ct dataset with manually annotated tumor lesions (psma-pet-ct-lesions). \url{https://doi.org/10.7937/r7ep-3x37} (2024), the Cancer Imaging Archive

\bibitem{kaji2024improvement}
Kaji, T., Osanai, K., Takahashi, A., Kinoshita, A., Satoh, D., Nakata, T., Tamaki, N.: Improvement of motion artifacts using dynamic whole-body 18{F}-{FDG} {PET}/{CT} imaging. Japanese Journal of Radiology  \textbf{42}(4),  374--381 (2024). \doi{https://doi.org/10.1007/s11604-023-01513-z}

\bibitem{kovacs2023addressing}
Kovacs, B., Netzer, N., Baumgartner, M., Schrader, A., Isensee, F., Wei{\ss}er, C., Wolf, I., G{\"o}rtz, M., Jaeger, P.F., Sch{\"u}tz, V., et~al.: Addressing image misalignments in multi-parametric prostate {MRI} for enhanced computer-aided diagnosis of prostate cancer. Scientific Reports  \textbf{13}(1),  19805 (2023). \doi{https://doi.org/10.1038/s41598-023-46747-z}

\bibitem{lodge2011effect}
Lodge, M.A., Mhlanga, J.C., Cho, S.Y., Wahl, R.L.: Effect of patient arm motion in whole-body {PET}/{CT}. Journal of Nuclear Medicine  \textbf{52}(12),  1891--1897 (2011). \doi{10.2967/jnumed.111.093583}

\bibitem{singh2023systematic}
Singh, P.: Systematic review of data-centric approaches in artificial intelligence and machine learning. Data Science and Management  \textbf{6}(3),  144--157 (2023)

\bibitem{zha2023data}
Zha, D., Bhat, Z.P., Lai, K.H., Yang, F., Jiang, Z., Zhong, S., Hu, X.: Data-centric artificial intelligence: A survey. arXiv preprint arXiv:2303.10158  (2023)

\end{thebibliography}

\end{document}